\documentclass[%
 reprint,
 superscriptaddress,
 bibnotes,
 amsmath,amssymb,
 aps,
 prl,
 floatfix
]{revtex4-2}

\usepackage[dvipdfmx]{graphicx}
\usepackage{dcolumn}
\usepackage{bm}
\usepackage{hyperref}
\usepackage{float}
\hypersetup{
  colorlinks   = true,
  urlcolor     = black,
  linkcolor    = black,
  citecolor   = blue
}
\usepackage{etoolbox}
\apptocmd{\sloppy}{\hbadness 10000\relax}{}{}

\begin{document}

\title{Out-of-phase Plasmon Excitations in the Trilayer Cuprate Bi$_2$Sr$_2$Ca$_2$Cu$_3$O$_{10+\delta}$}

\author{S. Nakata}
\affiliation{Department of Material Science, Graduate School of Science, University of Hyogo, Ako, Hyogo 678-1297, Japan}

\author{M. Bejas}
\affiliation{Facultad de Ciencias Exactas, Ingenier\'{i}a y Agrimensura and Instituto de F\'{i}sica de Rosario (UNR-CONICET), Avenida Pellegrini 250, 2000 Rosario, Argentina}

\author{J. Okamoto}
\affiliation{National Synchrotron Radiation Research Center, Hsinchu 300092, Taiwan}

\author{K. Yamamoto}
\affiliation{NanoTerasu Center, National Institutes for Quantum Science and Technology, Sendai, 980-8572, Japan}

\author{D. Shiga}
\affiliation{Institute of Multidisciplinary Research for Advanced Materials (IMRAM), Tohoku University, Sendai 980-8577, Japan}

\author{R. Takahashi}
\affiliation{Department of Material Science, Graduate School of Science, University of Hyogo, Ako, Hyogo 678-1297, Japan}

\author{H. Y. Huang}
\affiliation{National Synchrotron Radiation Research Center, Hsinchu 30076, Taiwan}

\author{H. Kumigashira}
\affiliation{Institute of Multidisciplinary Research for Advanced Materials (IMRAM), Tohoku University, Sendai 980-8577, Japan}

\author{H. Wadati}
\affiliation{Department of Material Science, Graduate School of Science, University of Hyogo, Ako, Hyogo 678-1297, Japan}

\author{J. Miyawaki}
\affiliation{NanoTerasu Center, National Institutes for Quantum Science and Technology, Sendai, 980-8572, Japan}

\author{S. Ishida}
\affiliation{National Institute of Advanced Industrial Science and Technology, Tsukuba, Ibaraki 305-8568, Japan}

\author{H. Eisaki}
\affiliation{National Institute of Advanced Industrial Science and Technology, Tsukuba, Ibaraki 305-8568, Japan}

\author{A. Fujimori}
\affiliation{National Synchrotron Radiation Research Center, Hsinchu 30076, Taiwan}
\affiliation{Center for Quantum Science and Technology and Department of Physics, National Tsing Hua University, Hsinchu 30013, Taiwan}
\affiliation{Department of Physics, University of Tokyo, Bunkyo-ku, Tokyo 113-0033, Japan}

\author{A. Greco}
\affiliation{Facultad de Ciencias Exactas, Ingenier\'{i}a y Agrimensura and Instituto de F\'{i}sica de Rosario (UNR-CONICET), Avenida Pellegrini 250, 2000 Rosario, Argentina}

\author{H. Yamase}
\email[]{yamase.hiroyuki@nims.go.jp}
\affiliation{Research Center for Materials Nanoarchitectonics (MANA),
National Institute for Materials Science (NIMS), Tsukuba 305-0047, Japan}

\author{D. J. Huang}
\affiliation{National Synchrotron Radiation Research Center, Hsinchu 30076, Taiwan}

\author{H. Suzuki}
\email[]{hakuto.suzuki@tohoku.ac.jp}
\affiliation{Frontier Research Institute for Interdisciplinary Sciences, Tohoku University, Sendai 980-8578, Japan}
\affiliation{Institute of Multidisciplinary Research for Advanced Materials (IMRAM), Tohoku University, Sendai 980-8577, Japan}
\date{\today}

\begin{abstract}
Within a homologous series of cuprate superconductors, variations in the stacking of CuO$_2$ layers influence the collective charge dynamics through the long-range Coulomb interactions. We use O $K$-edge resonant inelastic x-ray scattering to reveal plasmon excitations in the optimally-doped trilayer Bi$_2$Sr$_2$Ca$_2$Cu$_3$O$_{10+\delta}$. The observed plasmon exhibits nearly $q_z$-independent dispersion and a large excitation gap of approximately 300 meV. This mode is primarily ascribed to the $\omega_{-}$ mode, where the charge density on the outer CuO$_2$ sheets oscillates out of phase while the density in the inner sheet remains unaltered at $q_z=0$. The intensity of the acoustic $\omega_3$ mode is relatively weak and becomes vanishingly small near $(q_x, q_y)=(0, 0)$. This result highlights a qualitative change in the eigenmode of the dominant low-energy plasmon with the number of CuO$_2$ layers. 
\end{abstract}

\maketitle

High-temperature superconductivity in the cuprates emerges by introducing charge carriers into antiferromagnetic Mott insulators that contain the CuO$_2$ sheets \cite{Keimer.B_etal.Nature2015,Lee.P_etal.Rev.-Mod.-Phys.2006}. The proximity to the Mott insulating state commonly invokes theoretical description based on a single-band Hubbard model that describes the itinerant Cu 3$d_{x^2-y^2}$ holes on a square lattice and their on-site Coulomb repulsion. This approach describes the antiferromagnetic order in the parent compounds and the $d$-wave superconducting (SC) pairing in the doped compounds, arising from the virtual exchange of antiferromagnetic spin fluctuations \cite{Scalapino.D_etal.Rev.-Mod.-Phys.2012}.

Despite the success of spin-fluctuation theories in capturing key features of the cuprate superconductivity, it does not fully explain its material dependence. One of the major unanswered questions is the dependence of transition temperature ($T_c$) on the number of CuO$_2$ layers ($n$). Within a homologous series of cuprates, the $T_c$ monotonically increases with $n$ up to $n=3$ and decreases for $n \geq 4$ \cite{Iyo.A_etal.J.-Phys.-Soc.-Jpn.2007}. A theoretical proposal explaining this $n$ dependence of $T_c$ \cite{Leggett.A_etal.Phys.-Rev.-Lett.1999} suggests the importance of collective charge fluctuations arising from the long-range Coulomb interaction and its efficient
screening by the SC pairs, thereby enhancing $T_c$. Thus, it is crucial to reveal the $n$ dependence of charge fluctuations in the momentum space.

Recent advances in the energy resolution of resonant inelastic x-ray scattering (RIXS) \cite{Groot.F_etal.Nat.-Rev.-Methods-Primers2024} have enabled the measurement of elementary excitations relevant to superconductivity, including paramagnons \cite{Le-Tacon.M_etal.Nat.-Phys.2011,Peng.Y_etal.Nat.-Phys.2017}, phonons \cite{Devereaux.T_etal.Phys.-Rev.-X2016}, and plasmons \cite{Hepting.M_etal.Nature2018,Lin.J_etal.npj-Quantum-Mater.2020,Nag.A_etal.Phys.-Rev.-Lett.2020,Singh.A_etal.Phys.-Rev.-B2022,Hepting.M_etal.Phys.-Rev.-Lett.2022}. The plasmon excitations highlight the crucial role of the long-range Coulomb interaction in the low-energy charge dynamics of the cuprates, which is neglected in the Hubbard-model descriptions that incorporate only the local interactions.

\begin{figure}[htbp]
  \centering
  \includegraphics[angle = 0, width = 0.48\textwidth, clip=true]{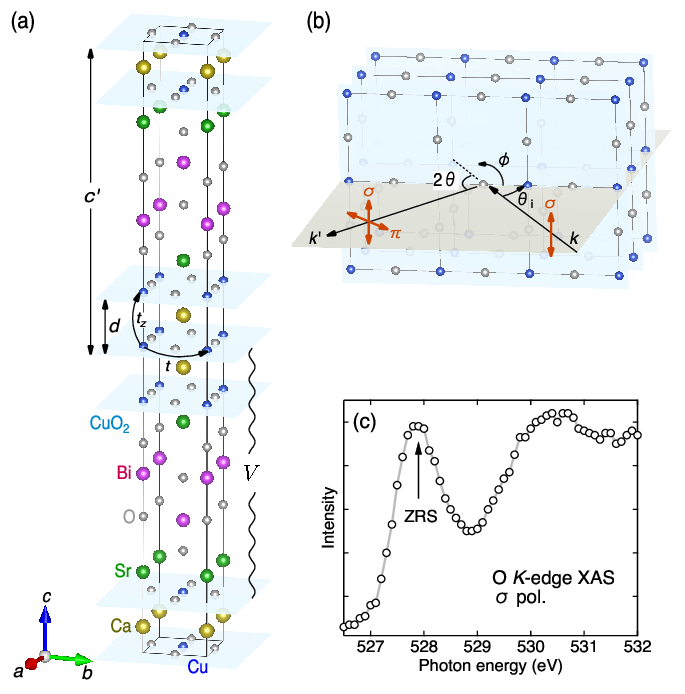}
  \caption{(a) Crystal structure of Bi$_2$Sr$_2$Ca$_2$Cu$_3$O$_{10+\delta}$ (Bi2223) \cite{Momma.K_etal.J.-Appl.-Cryst.2011}. The inter-trilayer distance $c^\prime$ is half of the crystallographic lattice constant $c=37.16$ \AA.   (b) Scattering geometry of the O $K$-edge resonant inelastic x-ray scattering (RIXS) experiment. The incident x-ray photons were $\sigma$-polarized, and the polarization of the scattered photons was not analyzed. (c) O $K$-edge x-ray absorption spectrum (XAS) measured at 24 K with the $\sigma$ polarization. The arrow indicates the absorption peak (527.9 eV) to the Zhang-Rice singlet (ZRS) states. \label{fig:1}}
\end{figure}

Plasmon dispersions identified in single-layer ($n=1$) compounds  La$_{2-x}$Ce$_x$CuO$_4$ \cite{Hepting.M_etal.Nature2018,Lin.J_etal.npj-Quantum-Mater.2020}, La$_{2-x}$Sr$_x$CuO$_4$ \cite{Nag.A_etal.Phys.-Rev.-Lett.2020,Singh.A_etal.Phys.-Rev.-B2022}, and Bi$_2$Sr$_{1.6}$La$_{0.4}$CuO$_{6+\delta}$ (Bi2201) \cite{Nag.A_etal.Phys.-Rev.-Lett.2020} are all acoustic-like, with a small energy gap in the limit $(q_x, q_y)\rightarrow (0, 0)$. In contrast, the infinite-layer ($n=\infty$) Sr$_{0.9}$La$_{0.1}$CuO$_2$ has a gapped plasmon branch \cite{Hepting.M_etal.Phys.-Rev.-Lett.2022}. This energy gap arises from interlayer electron hopping \cite{Greco.A_etal.Phys.-Rev.-B2016}, which brings the system closer to the three-dimensional electron gas with a gapped plasma frequency \cite{Fetter.A_etal.1971}. In both cases, the plasmons exhibit a large $q_z$ dispersion near $(q_x, q_y)=(0, 0)$ caused by an evolution from an optical mode at $q_z=0$ to acoustic-like modes at $q_z\ne 0$. In bilayer ($n=2$) compounds, the Coulomb interaction across the bilayers generates two plasmon branches at each $q_z$ \cite{Griffin.A_etal.Phys.-Rev.-B1989}, and the intra-bilayer electron hopping generates an energy gap in one of these branches \cite{Bejas.M_etal.Phys.-Rev.-B2024}. Experimental data of the bilayer Y$_{0.85}$Ca$_{0.15}$Ba$_2$Cu$_3$O$_7$ indicate that the gapped branch carries the dominant spectral weight \cite{Bejas.M_etal.Phys.-Rev.-B2024}. However, the lack of experimental data on cuprate compounds with a finite $n\geq 3$ hinders the understanding of the evolution of plasmon dispersions as a function of $n$, and the role of plasmons in the superconducting mechanism of the cuprates.

In this Letter, we present a RIXS investigation of the plasmon excitations in the optimally-doped trilayer cuprate Bi$_2$Sr$_2$Ca$_2$Cu$_3$O$_{10+\delta}$ (Bi2223), which shows the highest $T_c$ of 110 K among the Bi-based cuprates \cite{SM}. The observed plasmon branch exhibits a nearly two-dimensional dispersion with a significant gap of approximately 300 meV at the two-dimensional Brillouin zone center. Theoretical calculations of the charge susceptibility within the random phase approximation (RPA) predict three plasmon branches. We show that the observed dispersion is mainly described by the $\omega_{-}$ mode, which represents the out-of-phase oscillation of the charges on the outer CuO$_2$ sheets at $q_z=0$ and exhibits a weak $q_z$ dispersion, strikingly different from the plasmons observed in single-layer cuprates.

\begin{figure}[ht]
  \centering
  \includegraphics[angle = 0, width = 0.48\textwidth, clip=true]{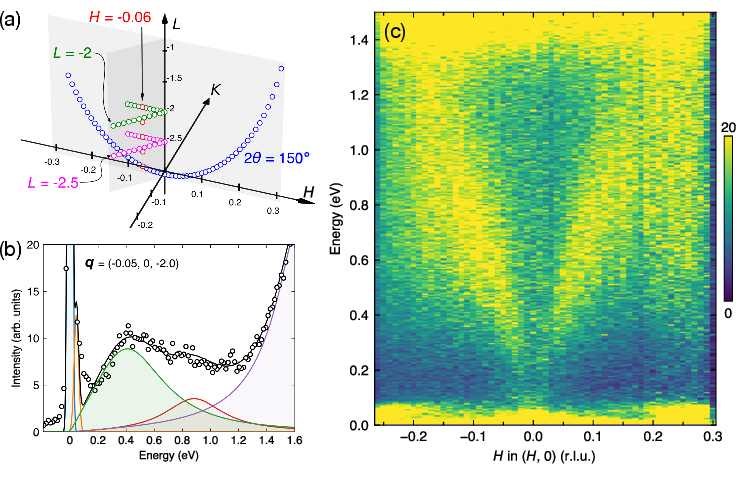}
  \caption{(a) Measurement paths in the three-dimensional ${\bm q}$ space. (b) Representative O $K$-edge RIXS spectrum of Bi2223 at ${\bm q}$ = (-0.05, 0, -2.0).   The spectrum includes the elastic line (blue), phonon (orange), plasmon (green), bimagnon (red), and $dd$ excitations (purple). (c) RIXS intensity map at 24 K along the ${\bm q}=(H, 0, L)$ direction taken with a maximal scattering angle of $2\theta=150^\circ$.} \label{fig:2}
\end{figure}

Figure \ref{fig:1}(a) shows the crystal structure of Bi2223 and the key parameters that determine the plasmon dispersion. The CuO$_2$ trilayers (light blue planes) are separated by spacer layers, with an inter-trilayer distance of $c^\prime=18.58$ \AA, which is half the crystallographic lattice constant $c=37.16$ \AA\ \cite{Fujii.T_etal.J.-Cryst.-Growth2001}. This large inter-trilayer distance suppresses electron hopping across the trilayers. In contrast, the interlayer distance within a trilayer, $d=3.30$ \AA, allows a finite interlayer hopping $t_z$. The $t_z$ combined with the long-range Coulomb interaction $V$ yields the gapped dispersion of the plasmon branch.

\begin{figure*}[ht]
  \centering
  \includegraphics[angle = 0, width = 1\textwidth, clip=true]{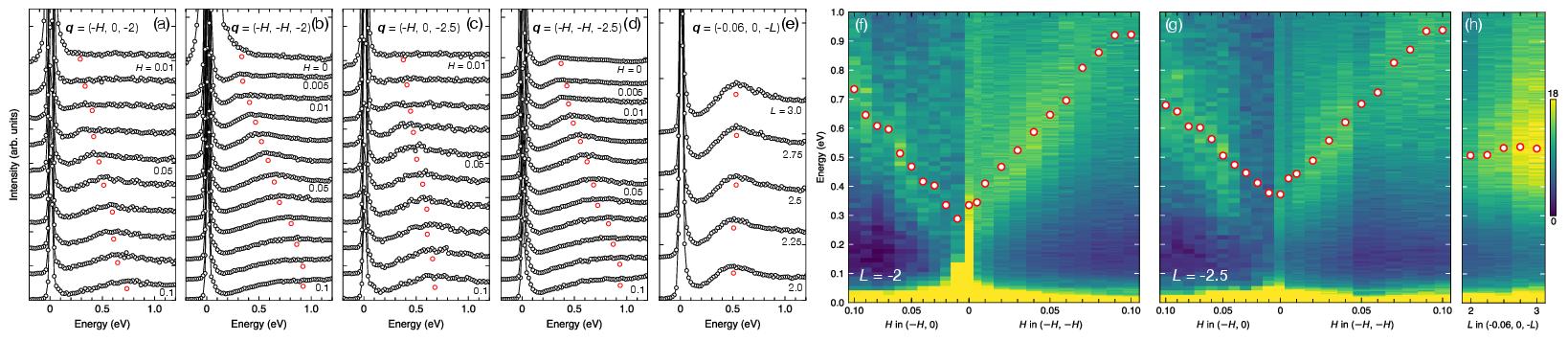}
  \caption{RIXS spectra at ${\bm q}=$($H$, 0, $L$) and ($H$, $H$, $L$), collected at fixed $L$ values (a,b) $L$ = -2 and (c,d) $L$ = -2.5. (e) RIXS spectra collected along the $L$ direction at ${\bm q}=(-0.06, 0, L)$. Red circles indicate the plasmon peak energies. (f)-(h) RIXS intensity maps corresponding to panels (a)-(e).}
 \label{fig:3}
\end{figure*}

To identify collective charge fluctuations in the hole-doped Bi2223, we employed O $K$-edge RIXS.  The scattering geometry of the RIXS experiment is illustrated in Fig. \ref{fig:1}(b). The incident x-ray photons were $\sigma$-polarized, and the outgoing photons with both $\sigma$ and $\pi$ polarizations were collected. Other experimental conditions are detailed in the Supplemental Material \cite{SM}.

The resonant transition to the itinerant O 2$p$ holes was determined by an O $K$-edge x-ray absorption spectrum (XAS) collected with $\sigma$-polarized photons. The XAS lineshape shown in Fig. \ref{fig:1}(c) reproduces previous results for both Pb-doped and Pb-free Bi2223 \cite{Karppinen.M_etal.Phys.-Rev.-B2003}. For RIXS measurements, we set the incident energy to 527.9 eV (indicated by the arrow), which corresponds to the transition to the Zhang-Rice singlet states formed by the hybridization between the Cu 3$d_{x^2-y^2}$ and O 2$p$ orbitals \cite{Zhang.F_etal.Phys.-Rev.-B1988}. This condition enhances the RIXS cross section of the charge scattering from the itinerant holes in the hole-doped cuprates \cite{Nag.A_etal.Phys.-Rev.-Lett.2020,Singh.A_etal.Phys.-Rev.-B2022}.

The investigated ${\bm q}$ paths in the reciprocal space are depicted in Fig. \ref{fig:2}(a). We first provide an overview of the charge fluctuations in Bi2223 by employing the largest scattering angle of $2\theta=150^\circ$ to maximize the momentum transfer $|{\bm q}|$. The in-plane component was scanned by rotating the sample angle (blue circles). Subsequently, we investigate the three-dimensional dispersions along straight lines in the ${\bm q}$ space by simultaneously tuning the $2\theta$ and sample angle $\theta_i$. Specifically, we explore the ${\bm q}=(H, 0, L)$ and $(H, H, L)$ paths on the fixed $L=-2$ (green) and $L=-2.5$ (pink) planes, as well as the ${\bm q}=(-0.06, 0, L)$ path in the range $-3.0<L<-2.0$ (red).

A representative O $K$-edge RIXS spectrum at ${\bm q}=(-0.05, 0, -2.0)$ is presented in Fig. \ref{fig:2}(b). In addition to the elastic line at $\omega=0$ eV (blue), it contains multiple elementary excitations in the spin-conserving channel, including phonon (orange), plasmon (green), bimagnon (red), and the tail of the $dd$ excitations (purple). In the subsequent discussion, we will focus on the dispersion of the plasmon branch.

A colormap of RIXS spectra at 24 K collected with a constant scattering angle of $2\theta=150^\circ$ is shown in Fig. \ref{fig:2}(c). A pronounced plasmon dispersion is observed, originating from the two-dimensional zone center and extending up to approximately 1 eV. While a sharp plasmon peak is well-defined for small $|H|$ values, the linewidth steadily increases with larger $|H|$, resulting in significant broadening for $|H|\gtrsim 0.2$. This broadening suggests the damping of the plasmon excitations into incoherent electron-hole pairs, likely due to the intersection of plasmons with the electron-hole continuum.
This Landau damping of plasmons was not identified in previous investigations \cite{Hepting.M_etal.Nature2018,Lin.J_etal.npj-Quantum-Mater.2020,Nag.A_etal.Phys.-Rev.-Lett.2020,Singh.A_etal.Phys.-Rev.-B2022,Hepting.M_etal.Phys.-Rev.-Lett.2022}, as the investigated in-plane ${\bm q}$ range was limited to $|H|\lesssim 0.2$.

We now examine the three-dimensional dispersion relation in more detail. In Figs. \ref{fig:3} (a)-(d), we show the RIXS spectra at 24 K at ${\bm q}=$($H$, 0, $L$) and ($H$, $H$, $L$), collected at fixed $L$ values  $L$ = -2 [(a,b)] and $L$ = -2.5 [(c,d)]. The plasmon peak positions are also shown as red circles. The corresponding intensity colormaps are presented in Figs. \ref{fig:3}(f) and (g), together with the peak positions. In contrast to the plasmons in single- \cite{Hepting.M_etal.Nature2018,Nag.A_etal.Phys.-Rev.-Lett.2020,Singh.A_etal.Phys.-Rev.-B2022} and infinite-layer \cite{Hepting.M_etal.Phys.-Rev.-Lett.2022} cuprates with significant $q_z$ dispersion, the dispersion of the observed branch in Bi2223 is nearly identical on the $L=-2$ and $L=-2.5$ planes, highlighting its two-dimensional character. Furthermore, it exhibits a large excitation gap of approximately 300 meV at the two-dimensional Brillouin zone center. 

To directly validate the small $q_z$ dispersion, we present RIXS spectra along the $L$ direction at ${\bm q}=(-0.06, 0, L)$ in Fig. \ref{fig:3}(e). The corresponding colormap is presented in Fig. \ref{fig:3}(h). At this in-plane momentum, the plasmon lineshape is almost independent of $L$, and the peak energy shows only a weak dispersion from 0.51 eV at $L=-2$ to 0.53 eV at $L=-3$. This energy difference at these equivalent ${\bm q}$ vectors shows that the periodicity of the plasmon intensity is determined by the inter-trilayer distance $c^\prime$, instead of the crystallographic lattice constant $c$ \cite{Hepting.M_etal.Nature2018}. This $q_z$ dispersion is significantly smaller than the plasmons in single- \cite{Hepting.M_etal.Nature2018,Lin.J_etal.npj-Quantum-Mater.2020,Nag.A_etal.Phys.-Rev.-Lett.2020,Singh.A_etal.Phys.-Rev.-B2022} and infinite-layer \cite{Hepting.M_etal.Phys.-Rev.-Lett.2022} cuprates. The small $q_z$ dispersion is a characteristic feature of the $\omega_{-}$ branch in trilayer cuprates, as shown below.

\begin{figure}[ht]
  \centering
  \includegraphics[angle = 0, width = 0.48\textwidth, clip=true]{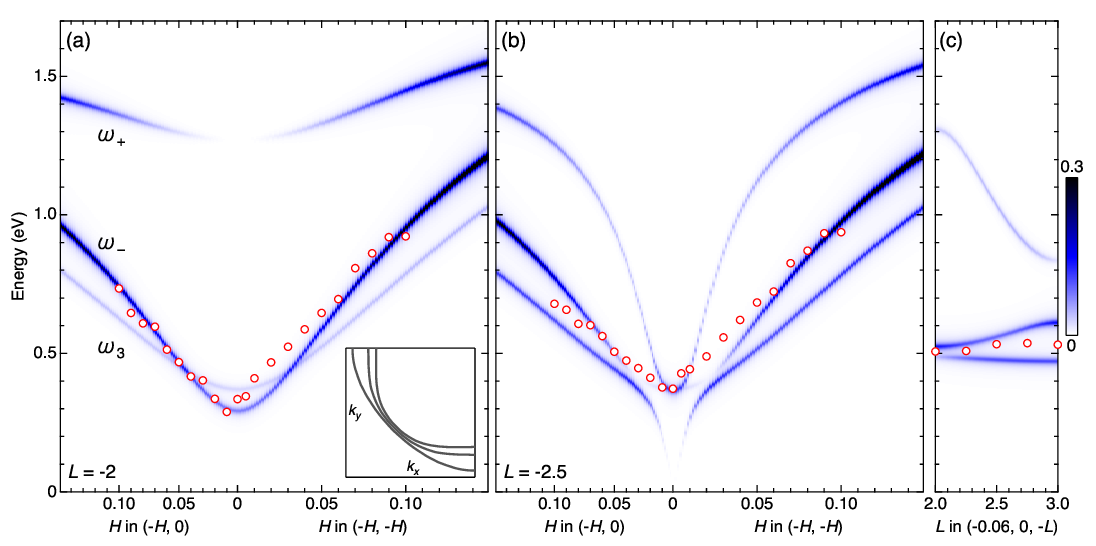}
  \caption{Dynamical charge susceptibility computed within the RPA. The intensity maps are generated along the experimental $\bm{q}$ paths. The notations of three plasmon branches ($\omega_{+}$, $\omega_{-}$, and $\omega_{3}$) follow those in Ref. \cite{Griffin.A_etal.Phys.-Rev.-B1989}. The experimental plasmon peak positions are overlaid as red open circles. The inset of panel (a) depicts the Fermi surfaces of the employed trilayer model \cite{SM}.}
  \label{fig:4}
\end{figure}

To quantitatively describe the observed plasmon dispersion, we have computed the dynamical charge susceptibility of  Bi2223 within the RPA. The theoretical trilayer model and the explicit form of the long-range Coulomb interactions are detailed in the Supplemental Material \cite{SM}. Our theoretical treatment is analogous to Ref. \cite{Griffin.A_etal.Phys.-Rev.-B1989}, but we adopt a tight-binding model with a finite intra-trilayer hopping $t_z$, which yields gapped plasmon branches at $(q_x, q_y)=(0, 0)$. The Fermi surfaces of the model [inset of Fig. \ref{fig:4}(a)] are in good agreement with angle-resolved photoemission data of optimally-doped Bi2223 \cite{Ideta.S_etal.Phys.-Rev.-Lett.2010,Kunisada.S_etal.Phys.-Rev.-Lett.2017}. We do not include the electron hopping between the outer CuO$_2$ sheets within a trilayer, which might become relevant only in the overdoped region \cite{Luo.X_etal.Nat.-Phys.2023}.

Figures \ref{fig:4}(a) and (b) show the intensity colormaps of computed charge susceptibility [-$\text{Im} \chi ({\bm q}, \omega)$] along the experimental ${\bm q}$ paths on the $L=-2$ and $-2.5$ planes, respectively. The experimental plasmon peak positions are also overlaid as red circles. Reflecting the presence of the three CuO$_2$ sheets, three plasmon branches appear within the RPA. The $\omega_{+}$ mode has strong $q_z$-dependence, resulting in different in-plane dispersions on the $L=-2$ and $-2.5$ planes. As its dominant spectral weight lies in the high-energy region ($\gtrsim 1.2$ eV) that falls within the tail of the $dd$ excitations (see Fig. \ref{fig:2}), it is not resolved in the present RIXS measurements. In the low-energy region ($\lesssim 1.2$ eV), the dominant spectral weight lies in the $\omega_{-}$ branch, although the $\omega_3$ mode also has sizable spectral weight, particularly at $L=-2.5$. Due to the finite $t_z$ and the long-range Coulomb interaction $V$, the dispersion of the $\omega_{-}$ mode becomes gapped at the two-dimensional zone center, providing an excellent description of the experimental plasmon dispersion. Note that the $\omega_3$ mode remains gapless at ${\bm q}=(0, 0, -2.5)$, but its spectral weight near ${\bm q}=(0, 0, -2.5)$ becomes vanishingly small. Concomitantly, the bottom of the $\omega_{+}$ mode starts to carry sizable spectral weight near ${\bm q}=(0, 0, -2.5)$, reproducing the experimental peak positions.

The intensity map along the ${\bm q}=(-0.06, 0, L)$ path, shown in Fig. \ref{fig:4}(c), reveals the distinct $q_z$ dependence of the three plasmon branches. The $\omega_{+}$ mode exhibits a strong $L$ dependence with its minimum at $L=-3$, similar to the plasmons in single- and infinite-layer cuprates. On the other hand, the energies of the $\omega_{-}$ and $\omega_{3}$ modes depend on $L$ only weakly, and the slight increase of the $\omega_{-}$ mode from $L=-2$ to $L=-3$ agrees with the experimental data.  

Here, we mention the linewidths of the plasmon excitations. The plasmon linewidth of the RIXS data is quite broad in the small in-plane $(q_x, q_y)$ region presented in Fig. \ref{fig:3}.  On the other hand, the calculated RPA charge susceptibility predicts the three sharp plasmon branches. This is because the RPA does not incorporate spectral broadening due to electron correlations, such as short-range antiferromagnetic correlations \cite{Prelovsek.P_etal.Phys.-Rev.-B1999}. In the single-layer case, this effect is discussed in the strong-coupling $t$-$J$-$V$ model \cite{Nag.A_etal.Phys.-Rev.-Res.2024}. It can be simulated phenomenologically by invoking a large broadening parameter comparable to the peak widths of the observed plasmons \cite{SM}.

Our RIXS data and theoretical analysis have revealed that intra-trilayer hopping in Bi2223 is responsible for the finite energy gap in the $\omega_{-}$ branch. Note that this mechanism differs from the energy gap generated by inter-(multi)layer hopping \cite{Hepting.M_etal.Phys.-Rev.-Lett.2022}, which generates the three-dimensionality of the electronic band structures. For instance, the $k_z$-dependent modulation of the Fermi surfaces in the overdoped La$_{2-x}$Sr$_x$CuO$_4$ ($x=0.22$) with an interlayer distance of 6.61 \AA\ shows that the magnitude of the interlayer hopping is 7 \% of the nearest-neighbor hopping \cite{Horio.M_etal.Phys.-Rev.-Lett.2018}. In contrast, the weak modulation observed in Tl$_2$Ba$_2$CuO$_{6+\delta}$, which has a long interlayer distance of 11.60 \AA, suggests that interlayer hopping is less than 1.5 \% of the nearest-neighbor hopping \cite{Horio.M_etal.Phys.-Rev.-Lett.2018}.
In Bi2223, the longer inter-trilayer distance, given by $c^\prime - 2d = 11.98$ \AA, likely leads to a negligibly small inter-trilayer electron hopping.

Now, we discuss the $n$ dependence of the plasmon branches in the absence of inter-multilayer hopping. In the single-layer case ($n=1$), only the $\omega_{+}$ mode exists, whose dispersion is acoustic-like, except for the optical branches appearing at integer $L$ values \cite{Fetter.A_etal.1971}. In the bilayer case ($n=2$), both $\omega_{+}$ and $\omega_{-}$ modes are present, and the intra-bilayer hopping generates an energy gap in one of them at $(q_x, q_y)=(0, 0)$. Ref. \cite{Bejas.M_etal.Phys.-Rev.-B2024} concluded that the $\omega_{+}$ mode carries the dominant spectral weight and explains the observed dispersion in Y$_{0.85}$Ca$_{0.15}$Ba$_2$Cu$_3$O$_7$, although recent analyses suggested that the observed dispersion can be explained by the $\omega_-$ mode \cite{Yamase.H_etal.2024,Sellati.N_etal.2024}. In the trilayer case ($n=3$), the three modes ($\omega_{+}$, $\omega_{-}$, and $\omega_{3}$) emerge, and we have shown that the $\omega_3$ and $\omega_{-}$ modes exhibit a weak $q_z$ dependence. The observed plasmon in Bi2223 is primarily ascribed to the $\omega_{-}$ branch with the dominant spectral weight, supplemented by contributions from the $\omega_3$ mode at $L=-2$ and from the $\omega_3$ and $\omega_+$ modes at $L=-2.5$.

These results highlight the qualitative change in the eigenmode of the dominant charge fluctuations as a function of $n$ in cuprates. For simplicity, we describe the eigenmodes of the $\omega_{+}$, $\omega_{-}$, and $\omega_{3}$ modes at $q_z=0$. The $\omega_{+}$ mode, present in any $n$, is the in-phase charge oscillation among the multilayers. The $\omega_{-}$ mode, present in $n\geq 2$, is an out-of-phase charge oscillations within the multilayers. In the trilayer case, the charge density on the outer two sheets oscillates out of phase, while the density on the inner sheet remains unaltered. The $\omega_{3}$ mode appears in $n\geq 3$. In the trilayer case, the charge density on the outer and inner sheets oscillates out of phase, with the larger amplitude in the inner sheet. While the plasmons in the single- and infinite-layer cuprates \cite{Hepting.M_etal.Nature2018,Lin.J_etal.npj-Quantum-Mater.2020,Nag.A_etal.Phys.-Rev.-Lett.2020,Singh.A_etal.Phys.-Rev.-B2022,Hepting.M_etal.Phys.-Rev.-Lett.2022} originate from the $\omega_{+}$ mode, our results on the trilayer Bi2223 provide the first identification of the $\omega_{-}$ plasmon mode in cuprates. Our results also call for more comprehensive data in bilayer cuprates \cite{Bejas.M_etal.Phys.-Rev.-B2024}, for the precise assignment of the eigenmode.

A natural question is how the evolution of plasmons with $n$ affects the SC properties in cuprates. We note that the lineshape of the RIXS spectrum at ${\bm q}=(-0.1, 0, -2)$ at 300 K ($>T_c$) remains almost identical to that at 25 K \cite{SM}. This indicates that the reconstruction in the dynamical charge susceptibility in the mid-infrared region across $T_c$, as theoretically proposed \cite{Leggett.A_etal.Phys.-Rev.-Lett.1999}, was not detected at this ${\bm q}$. In Bi2223,  the bottom energy of the plasmons, $\omega \sim$ 300 meV, is significantly larger than 2$\Delta$, where $\Delta$ is the amplitude of its $d$-wave SC order parameter in the three Fermi surfaces \cite{Ideta.S_etal.Phys.-Rev.-Lett.2010,Kunisada.S_etal.Phys.-Rev.-Lett.2017,Ideta.S_etal.Phys.-Rev.-Lett.2021}. This is not the case in single-layer cuprates with acoustic-like plasmon dispersion. Future investigations of the detailed temperature dependence of plasmon lineshape across $T_c$, both in single- and multilayer cuprates, will shed light on the role of plasmons in the superconducting mechanism.

In conclusion, we have conducted an O $K$-edge RIXS study of plasmon excitations in the optimally doped trilayer cuprate Bi2223. We have identified nearly two-dimensional plasmon excitations, which exhibit a large excitation gap of approximately 300 meV at $(q_x, q_y)= (0, 0)$ generated by the long-range Coulomb interactions and the intra-trilayer hopping. We attribute this mode primarily to the $\omega_{-}$ mode, in contrast to the $\omega_{+}$ mode identified in single- and infinite-layer cuprates. These results suggest a qualitative change in the eigenmode of the dominant plasmon with the number of CuO$_2$ layers in cuprate superconductors.

\textit{Acknowledgements.}---
We thank T. Tohyama for enlightening discussions.
The project was supported by Grants-in-Aid for Scientific Research from JSPS (KAKENHI) (No.~JP19H05823,  JP22K13994). M.B. is indebted to MANA Short-Term Invitation Program and warm hospitality in NIMS; he was also partially supported by JSPS KAKENHI (Grant No.~JP20H01856).  A.F. was supported by JSPS KAKENHI (Grant No. JP22K03535), NSTC Taiwan (Grant No. NSTC 113-2112-M-007-033), and the Yushan Fellow Program and the Center for Quantum Science and Technology within the framework of the Higher Education Sprout Project under the MOE of Taiwan. H.Y. was supported by JSPS KAKENHI (Grant No.~JP20H01856) and the World Premier International  Research Center Initiative (WPI), MEXT, Japan. Part of the theoretical results was obtained by using the facilities of the CCT-Rosario Computational Center, a member of the High Performance Computing National System (SNCAD, MincyT-Argentina).

\bibliography{Bi2223_plasmon}

\end{document}


\title{Supplemental Material for\\ Out-of-phase Plasmon Excitations in the Trilayer Cuprate Bi$_2$Sr$_2$Ca$_2$Cu$_3$O$_{10+\delta}$}

\author{S. Nakata}
\affiliation{Department of Material Science, Graduate School of Science, University of Hyogo, Ako, Hyogo 678-1297, Japan}

\author{M. Bejas}
\affiliation{Facultad de Ciencias Exactas, Ingenier\'{i}a y Agrimensura and Instituto de F\'{i}sica de Rosario (UNR-CONICET), Avenida Pellegrini 250, 2000 Rosario, Argentina}

\author{J. Okamoto}
\affiliation{National Synchrotron Radiation Research Center, Hsinchu 300092, Taiwan}

\author{K. Yamamoto}
\affiliation{NanoTerasu Center, National Institutes for Quantum Science and Technology, Sendai, 980-8572, Japan}

\author{D. Shiga}
\affiliation{Institute of Multidisciplinary Research for Advanced Materials (IMRAM), Tohoku University, Sendai 980-8577, Japan}

\author{R. Takahashi}
\affiliation{Department of Material Science, Graduate School of Science, University of Hyogo, Ako, Hyogo 678-1297, Japan}

\author{H. Y. Huang}
\affiliation{National Synchrotron Radiation Research Center, Hsinchu 300092, Taiwan}

\author{H. Kumigashira}
\affiliation{Institute of Multidisciplinary Research for Advanced Materials (IMRAM), Tohoku University, Sendai 980-8577, Japan}

\author{H. Wadati}
\affiliation{Department of Material Science, Graduate School of Science, University of Hyogo, Ako, Hyogo 678-1297, Japan}

\author{J. Miyawaki}
\affiliation{NanoTerasu Center, National Institutes for Quantum Science and Technology, Sendai, 980-8572, Japan}

\author{S. Ishida}
\affiliation{National Institute of Advanced Industrial Science and Technology, Tsukuba, Ibaraki 305-8568, Japan}

\author{H. Eisaki}
\affiliation{National Institute of Advanced Industrial Science and Technology, Tsukuba, Ibaraki 305-8568, Japan}

\author{A. Fujimori}
\affiliation{National Synchrotron Radiation Research Center, Hsinchu 300092, Taiwan}
\affiliation{Center for Quantum Science and Technology and Department of Physics, National Tsing Hua University, Hsinchu 30013, Taiwan}
\affiliation{Department of Physics, University of Tokyo, Bunkyo-ku, Tokyo 113-0033, Japan}

\author{A. Greco}
\affiliation{Facultad de Ciencias Exactas, Ingenier\'{i}a y Agrimensura and Instituto de F\'{i}sica de Rosario (UNR-CONICET), Avenida Pellegrini 250, 2000 Rosario, Argentina}

\author{H. Yamase}
\affiliation{Research Center for Materials Nanoarchitectonics (MANA),
National Institute for Materials Science (NIMS), Tsukuba 305-0047, Japan}

\author{D. J. Huang}
\affiliation{National Synchrotron Radiation Research Center, Hsinchu 300092, Taiwan}

\author{H. Suzuki}
\affiliation{Frontier Research Institute for Interdisciplinary Sciences, Tohoku University, Sendai 980-8578, Japan}
\affiliation{Institute of Multidisciplinary Research for Advanced Materials (IMRAM), Tohoku University, Sendai 980-8577, Japan}

\maketitle

\section{\NoCaseChange{Characterization of Bi$_2$Sr$_2$Ca$_2$Cu$_3$O$_{10+\delta}$ single crystals}}

\begin{figure}[hbt]
  \centering
  \includegraphics[width = 0.7\textwidth, clip=true]{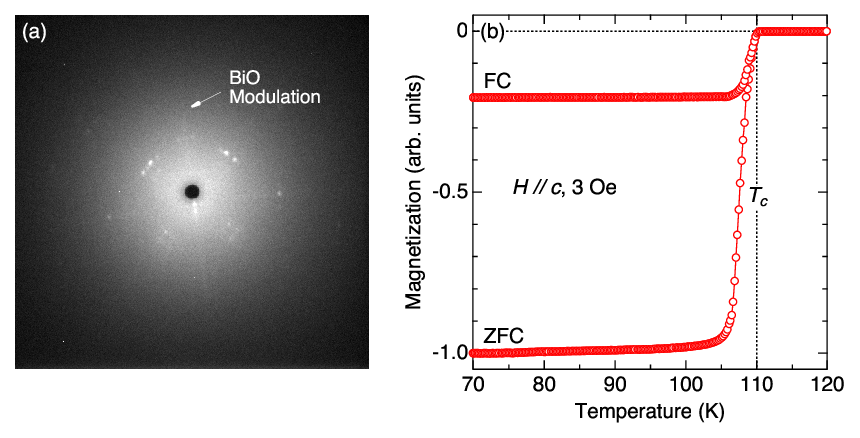}
  \caption{\label{fig:S1} Laue diffraction pattern for a Bi2223 crystal. (b) Magnetization curves for the optimally-doped Bi2223 ($T_c = 110$ K) measured with the field cooling (FC) and zero-field cooling (ZFC) methods. A magnetic field of 3 Oe was applied along the crystallographic $c$ axis.}
\end{figure}

The orientation of optimally-doped Bi$_2$Sr$_2$Ca$_2$Cu$_3$O$_{10+\delta}$ (Bi2223) single crystal measured in the RIXS experiment was determined from the Laue diffraction pattern [Fig. \ref{fig:S1}(a)]. The straight line corresponds to charge modulation in the BiO planes along the ${\bm q}=(H, H, L)$ direction. The temperature dependence of magnetic susceptibility is shown in Fig. \ref{fig:S1}(b). The data were collected with an applied magnetic field of 3 Oe along the crystallographic $c$ axis by field cooling (FC) and zero-field cooling (ZFC) methods. The superconducting transition temperature ($T_c = 110$ K) was defined as the onset temperature of the Meissner diamagnetic signal.

\clearpage
\section{\NoCaseChange{Experimental conditions for RIXS measurements}}
O $K$-edge RIXS spectra at 24 K were collected using the AGS-AGM spectrometer \cite{Singh.A_etal.J.-Synchrotron-Rad.2021} at Beamline 41A of Taiwan Photon Source.  The continuous rotation of the detector arm \cite{Singh.A_etal.J.-Synchrotron-Rad.2021} permits ${\bm q}$-space scans with fixed $H$ or $L$ values. The total energy resolution was set to 22 meV, as estimated from the full-width at half maximum (FWHM) of the nonresonant spectrum from carbon tape. In addition, a RIXS spectrum at 300 K in a wide energy region was collected using the 2D-RIXS spectrometer \cite{Miyawaki.J_etal.J.-Phys.:-Conf.-Ser.2022} at BL02U of NanoTerasu.  Before installation in the measurement chamber, Bi2223 single crystals were cleaved \textit{ex situ} to prepare a fresh surface parallel to the CuO$_2$ planes. In the main text, we use the pseudotetragonal notation with lattice constants $a=b=3.85$  \AA\ and $c=37.16$ \AA\ \cite{Fujii.T_etal.J.-Cryst.-Growth2001}. The momentum transfer ${\bm q}=(H, K, L)$ is expressed in the reciprocal lattice units $(2\pi/a, 2\pi/b, 2\pi/c)$.

\section{\NoCaseChange{RIXS spectrum at 300 K in a wide energy region}}

\begin{figure}[hbt]
  \centering
  \includegraphics[angle = 0, width = 0.5\textwidth, clip=true]{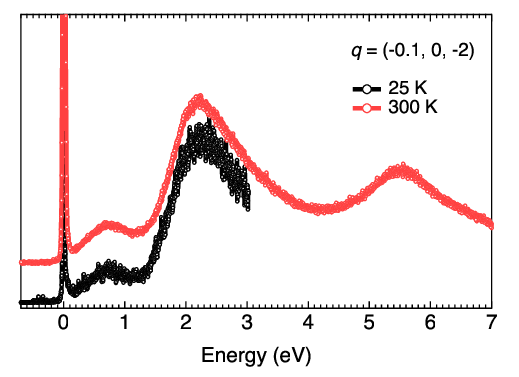}
  \caption{\label{wide} RIXS spectrum at ${\bm q}=(-0.1, 0, -2)$ at 300 K in a wide energy region (red). The corresponding spectrum at 25 K (black) is also shown for comparison.}
\end{figure}

Figure \ref{wide} presents the RIXS spectrum at ${\bm q}=(-0.1, 0, -2)$ in a wider energy region at 300 K. The corresponding spectrum at 25 K is also shown for comparison. We identify the charge-transfer excitation at $\sim$ 5.5 eV. We also note that the lineshapes of the plasmon excitation are almost identical between 25 K and 300 K. This indicates that the superconducting transition at $T_c = 110$ K does not have an appreciable feedback effect on the plasmon excitations.

\section{\NoCaseChange{Trilayer model and RPA calculations of charge susceptibility}}

The bare Green's function for the trilayer model is expressed as

\begin{align}
 G_0^{-1}({\bf k}, {\rm i}\nu_n) =& 
 \left[
 \begin{matrix}
   {\rm i}\nu_n - \varepsilon_{\bf k} & -\varepsilon^{\perp}_{\bf k} e^{{\rm i} k_z d} & 0 \\
   -\varepsilon^{\perp}_{\bf k} e^{-{\rm i} k_z d} & {\rm i}\nu_n - \varepsilon_{\bf k} & -\varepsilon^{\perp}_{\bf k} e^{{\rm i} k_z d} \\
   0 & -\varepsilon^{\perp}_{\bf k} e^{-{\rm i} k_z d} & {\rm i}\nu_n - \varepsilon_{\bf k}
 \end{matrix}
 \right]
\; ,
 \label{eq:G0m1}
\end{align}
\noindent where $\varepsilon_{\bf k} = -2t (\cos k_x + \cos k_y)
-4t' \cos k_x \cos k_y
-2t'' (\cos 2k_x  + \cos 2k_y) - \mu$; 
$\varepsilon^{\perp}_{\bf k} = -t_z (\cos k_x - \cos k_y)^2 - t_{z0}$; $k_x$ and $k_y$ are in units of $1/a$; $k_z$ is in units of $1/c'$; $a$ is the lattice constant along the $x$ and $y$ directions; $c'$ is half of the lattice constant $c$ since the unit cell contains two slabs of trilayers; the three layers are placed with equal distance $d$ in the trilayer slab; see Fig. 1.  
Defining
\begin{align}
 \varepsilon_1 &= \varepsilon_{\bf k} + \sqrt{2} \, \varepsilon^{\perp}_{\bf k} \; , \\
 \varepsilon_2 &= \varepsilon_{\bf k}  \; , \\
 \varepsilon_3 &= \varepsilon_{\bf k} - \sqrt{2} \, \varepsilon^{\perp}_{\bf k} \; , 
\end{align}
\noindent we obtain
\begin{align}
 G_0({\bf k}, {\rm i}\nu_n) =& \frac{1}{D}
 \left[
 \begin{matrix}
   ({\rm i}\nu_n - \varepsilon_{\bf k})^2 - (\varepsilon^{\perp}_{\bf k})^2 & ({\rm i}\nu_n - \varepsilon_{\bf k})\varepsilon^{\perp}_{\bf k} e^{{\rm i} k_z d} & (\varepsilon^{\perp}_{\bf k})^2 e^{{\rm i} 2k_z d} \\
    ({\rm i}\nu_n - \varepsilon_{\bf k})\varepsilon^{\perp}_{\bf k} e^{-{\rm i} k_z d} & ({\rm i}\nu_n - \varepsilon_{\bf k})^2 & ({\rm i}\nu_n - \varepsilon_{\bf k})\varepsilon^{\perp}_{\bf k} e^{{\rm i} k_z d} \\
    (\varepsilon^{\perp}_{\bf k})^2 e^{-{\rm i} 2k_z d} & ({\rm i}\nu_n - \varepsilon_{\bf k})\varepsilon^{\perp}_{\bf k} e^{-{\rm i} k_z d} & ({\rm i}\nu_n - \varepsilon_{\bf k})^2 - (\varepsilon^{\perp}_{\bf k})^2
 \end{matrix}
 \right]
\; ,
 \label{eq:G0}
\end{align}
\noindent with
$D = ({\rm i}\nu_n - \varepsilon_1) ({\rm i}\nu_n - \varepsilon_2)({\rm i}\nu_n - \varepsilon_3)$.

The non-interacting susceptibility matrix $\chi_0({\bf q}, {\rm i}\omega_n)$ is given by 
\begin{align}
 \chi_{0,ij}({\bf q}, {\rm i}\omega_n) &= \frac{1}{N_{x,y}N_z}\sum_{{\bf k}, {\rm i}\nu_n}
                              G_{0,ij}({\bf k}, {\rm i}\nu_n) \;
                              G_{0,ji}({\bf k} + {\bf q}, {\rm i}\nu_n + {\rm i}\omega_n) \; ,
 \label{eq:chi_0}
\end{align}
\noindent where $N_{x,y}$ ($N_z$) is the number of points in the ${\bf k_{\parallel}}$ ($k_z$) summation.  Note that the $k_z$ dependence only appears through the phases in the off-diagonal elements and the energies do not depend on $k_z$. Therefore we can extract the phase factor by introducing a quantity $d_{ij}$ such that $d_{ij}=0$ for $i=j$, $d_{ij}=d$ for $(i,j) = (1,2)$, $(2,3)$, $d_{ij}=2d$ for $(i,j)=(1,3)$, and $d_{ji}=-d_{ij}$.  
Consequently, Eq.(\ref{eq:chi_0}) is written as
\begin{align}
 \chi_{0,ij}({\bf q}, {\rm i}\omega_n) &= \frac{1}{N_{x,y}N_z}\sum_{{\bf k_{\parallel}}, k_z, {\rm i}\nu_n}
                              \tilde{G}_{0,ij}({\bf k}, {\rm i}\nu_n) \;
                              \tilde{G}_{0,ji}({\bf k} + {\bf q}, {\rm i}\nu_n + {\rm i}\omega_n)
                              e^{{\rm i} k_z d_{ij}} e^{-{\rm i} (k_z + q_z) d_{ij}}
                              \; ,\\
                              &= \frac{e^{-{\rm i} q_z d_{ij}}}{N_{x,y}} \sum_{\bf k_{\parallel}, {\rm i}\nu_n}
                              \tilde{G}_{0,ij}({\bf k}, {\rm i}\nu_n) \;
                              \tilde{G}_{0,ji}({\bf k} + {\bf q}, {\rm i}\nu_n + {\rm i}\omega_n)
                              \; . 
 \label{eq:chi_0p}
\end{align}

Here $\tilde{G}_{0,ij}({\bf k}, {\rm i}\nu_n)$ is equal to $G_{0,ij}({\bf k}, {\rm i}\nu_n)$ without the phase factor. Then the matrix $\chi_0$ for the stack of trilayers is equal to the $\chi_0$ for only three layers with open boundary conditions, with the addition of phase factors $e^{\pm {\rm i} q_z d}$ and $e^{\pm {\rm i} 2q_z d}$ in the off-diagonal elements.

Following the long-range Coulomb interaction obtained in Ref.~\cite{Griffin.A_etal.Phys.-Rev.-B1989}, we employ 
\begin{align}
 V_{\bf q} =& 
 \left[
 \begin{matrix}
   V_1 & V_2 & V_3 \\
   V^*_2 & V_1 & V_2 \\
   V^*_3 & V^*_2 & V_1
 \end{matrix}
 \right]
\; ,
 \label{eq:Vq}
\end{align}
\noindent with
\begin{align}
  V_1 &= \frac{V_c}{q_{\parallel} a} \frac{\sinh(q_{\parallel} c')}
                                       {\cosh(q_{\parallel} c') - \cos(q_z c')} \; , \\
  V_2 &= \frac{V_c}{q_{\parallel} a}
        \frac{\sinh[q_{\parallel}(c'-d)] + e^{-{\rm i}q_z c'} \sinh(q_{\parallel} d)}
             {\cosh(q_{\parallel} c') - \cos(q_z c')} \,
             e^{{\rm i} q_z d}  \; , \\
  V_3 &= \frac{V_c}{q_{\parallel} a}
        \frac{\sinh[q_{\parallel}(c' - 2d)] + e^{-{\rm i}q_z c'} \sinh(q_{\parallel} 2d)}
             {\cosh(q_{\parallel} c') - \cos (q_z c')} \,
             e^{{\rm i} q_z 2d} \; ,
\end{align}

\noindent and $q_{\parallel} = \sqrt{q^2_x + q^2_y}$. The RPA susceptibility is then computed as 
\begin{align}
 \chi = (1 - \chi_0 V_{\bf q})^{-1} \; \chi_0 \; .
 \label{eq:chi}
\end{align}

Figure 4 in the main text presents $-{\rm Im} \chi_{\rm tot} = -{\rm Im} \sum_{i,j} \chi_{ij}$, where we use a doping $\delta = 0.18$, temperature $T = 0$, $t' = -0.26$, $t'' = 0.13$, $t_z = 0.07$, $t_{z0} = 0.1$, ${\rm i}\omega_n = \omega + {\rm i}\Gamma$, $\Gamma = 0.1$, $V_c = 280$; the energy unit is $t = 0.1$ eV; the lattice parameters are  $a = 3.85$ \r{A}, $c' = 18.58$ \r{A}, and $d = 0.86 a = 0.178 c'$.

\section{\NoCaseChange{Charge susceptibility with larger broadening}}
The broadening of plasmons is controlled by $\Gamma$ in the present theory. In principle, $\Gamma$ should be infinitesimally small, but we may invoke a finite value, e.g., $\Gamma = 0.1$, to mimic intrinsic broadening due to incoherent effects from electronic correlations \cite{Prelovsek.P_etal.Phys.-Rev.-B1999,Nag.A_etal.Phys.-Rev.-Res.2024}.  Phenomenologically, the choice of $\Gamma$ can be arbitrary. When we respect the broadening closer to the experimental linewidths [see the green curve in Fig.~2(b)], we may choose a larger $\Gamma=1.0$. In this case, we obtain Fig.~S3, which should be compared with Fig.~4 of the main text, where a smaller $\Gamma=0.1$ is employed. The $\omega_{3}$ and $\omega_-$ branches are combined into 
one broader branch, which well reproduces the experimental plasmon lineshapes.  

\begin{figure}[hbt]
  \centering
  \includegraphics[width = 0.8\textwidth, clip=true]{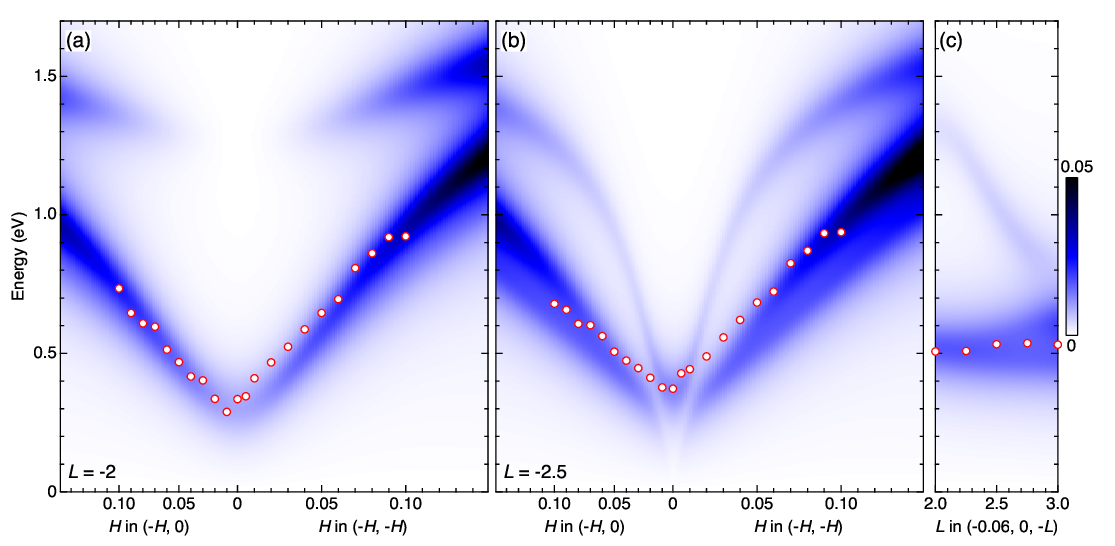}
  \caption{\label{fig:G10} Dynamical charge susceptibility computed for a large broadening parameter $\Gamma=1.0$.}
\end{figure}

\bibliography{Bi2223_plasmon}